\documentclass[journal]{IEEEtran}
\usepackage[utf8x]{inputenc}
\usepackage{amsmath}
\usepackage{graphicx}
\usepackage{epstopdf}
\usepackage{tikz,pgfplots}
\usetikzlibrary{shapes,decorations}
\usepgfplotslibrary{fillbetween}
\pgfplotsset{compat=newest}
\usepgfplotslibrary{groupplots}
\usepgfplotslibrary{dateplot}
\usepackage{mathrsfs}
\usepackage{eso-pic}
\usepackage{amssymb}
\usepackage{mathrsfs}
\usepackage{hyperref}
\usepackage{cancel}
 
\usepackage{ulem}
\newcommand{\llangle}[1][]{\savebox{\@brx}{\(\m@th{#1\langle}\)}%
  \mathopen{\copy\@brx\kern-0.5\wd\@brx\usebox{\@brx}}}
\newcommand{\rrangle}[1][]{\savebox{\@brx}{\(\m@th{#1\rangle}\)}%
  \mathclose{\copy\@brx\kern-0.5\wd\@brx\usebox{\@brx}}}
\makeatother

\newcommand{\Vext}{V_{e}}
\newcommand{\tOmega}{V}
\newcommand{\ptOmega}{S}
\newcommand{\tSigma}{  \Sigma}

\newcommand{\dV}{d^3\mathbf{r}}
\newcommand{\dS}{d^2\mathbf{r}}

\newcommand{\Surf}{S}
\newcommand{\sizep}{\xi}

\newcommand{\Jlo}{{\bf j}^{\parallel}}

\newcommand{\gammalo}[1]{\chi_{#1}^\parallel}

\newcommand{\Jto}{{\bf j}^{\perp}}
\newcommand{\gammato}[1]{\chi_{#1}^\perp}

\newcommand{\rt}{{\mathbf{r}}}
\newcommand{\rbt}{{\mathbf{r}}}
\newcommand{\rp}{\left(  {\mathbf{r}} \right)}
\newcommand{\rpp}{\left( {\mathbf{r}}' \right)}
\newcommand{\surfcond}{\Sigma}

\newcommand{\n}{\hat{\mathbf{n}}}

\newcommand{\tnabla}{{\nabla}}

\newcommand{\PM}[1]{\mathbf{M}_{#1}}

\newcommand{\PE}[1]{\mathbf{P}_{#1}}

\newcommand{\ca}{\hat{\mathbf{e}}}
\newcommand{\diropt}{\hat{\mathbf{p}}_\text{opt}}
\newcommand{\diroptOrt}{\hat{\mathbf{m}}_\text{opt}}

\newcommand{\jlopt}{\mathbf{j}}
\newcommand{\jtopt}{\mathbf{j}}

\newcommand{\QMSuni}{\hat{\mathbf{r}} \times {\bf c}}

\newcommand{\TensorM}{\overleftrightarrow{\boldsymbol{\gamma}}_m}
\newcommand{\TensorE}{\overleftrightarrow{\boldsymbol{\gamma}}_e}

\begin{document}

\title{Lower Bounds to the Q factor  \\ of Electrically Small Resonators \\
through Quasistatic Modal Expansion }

\author{\IEEEauthorblockN{Mariano Pascale, Sander A. Mann, Dimitrios C. Tzarouchis, Giovanni Miano, Andrea Alù, Carlo Forestiere}
\IEEEcompsocitemizethanks{
\IEEEcompsocthanksitem M. Pascale is with ICFO-Institut de Ciencies Fotoniques, The Barcelona Institute of Science and Technology, Castelldefels (Barcelona) 08860, Spain
\IEEEcompsocthanksitem M. Pascale, C. Forestiere and G. Miano are with the Department of Electrical Engineering and Information Technology, Universit\`{a} degli Studi di Napoli Federico II, via Claudio 21,  Napoli, 80125, Italy
\IEEEcompsocthanksitem M. Pascale, S. A. Mann and A. Alù are with the Photonics Initiative, Advanced Science Research Center, City University of New York, New York, New York 10031, U.S.A.
\IEEEcompsocthanksitem A. Alù is with the Physics Program, Graduate Center of the City University of New York, New York, New York 10016, U.S.A.
\IEEEcompsocthanksitem D. C. Tzarouchis is with the Department of Electrical and System Engineering, University of Pennsylvania, Philadelphia, PA 19104 U.S.A.}}

\maketitle

\begin{abstract}
The problem of finding the optimal current distribution supported by small radiators yielding the minimum quality (Q) factor is a fundamental problem in electromagnetism.
Q factor bounds constrain the maximum operational bandwidth of devices including antennas, metamaterials, and nanoresonators, and have been featured in seminal papers in the past decades. 
Here, we determine the lower bounds of Q factors of small-size plasmonic and high-permittivity dielectric resonators, which are characterized by quasi-electrostatic and quasi-magnetostatic natural modes, respectively. We expand 
the induced current density field in the resonator in terms of these modes, leading to closed-form analytical expressions for the electric and magnetic polarizability tensors, whose largest eigenvalue is directly linked to the minimum Q factor. Our results allow also to determine in closed form the corresponding optimal current density field. In particular, when the resonator exhibits two orthogonal reflection symmetries the minimum Q factor can be simply obtained from the Q factors of the single current modes with non-vanishing dipole moments aligned along the major axis of the resonator. Overall, our results open exciting opportunities in the context of nano-optics and metamaterials, facilitating the analysis and design of optimally shaped resonators for enhanced and tailored light-matter interactions.
\end{abstract}

\begin{IEEEkeywords}
 scattering,
eigenvalues and eigenfunctions, resonances, Q factors, plasmons, dielectric resonators
\end{IEEEkeywords}

\newif\iffig
\figtrue
\section{Introduction}
Chu's limit~\cite{chu_physical_1948} determines the minimum radiation quality (Q) factor of electrically small antennas. This limit applies to both self-resonant and non self-resonant antennas, provided that a convenient tuning network is used for the latter. The minimum Q factor is associated with an {\it optimal current} distribution supported by the antenna. The search for such lower bounds originated with the works of Chu~\cite{chu_physical_1948}, Wheeler~\cite{wheeler_fundamental_1947}, and Harrington \cite{harrington_effect_1960}, and several techniques have been proposed over the years by many contributors including Collin and Rothschild~\cite{collin_evaluation_1964}, and McLean~\cite{mclean_re-examination_1996}. Thal~\cite{thal_new_2006}, by restricting the sources to
only the electric surface currents producing nonzero fields within the volume of the antenna, he arrived at stricter bounds than his predecessors. Then, in a series of contributions, starting in Ref. \cite{gustafsson_physical_2007}, Gustafsson and coworkers provided shape-dependent bounds on the radiator's minimum Q, linking it to the available volume in which the search of the optimal current is constrained. They also reduced the variational problem of finding the minimum Q of antennas to determine the largest eigenvalues of the  polarizability tensor. In subsequent years, Gustafsson and co-workers refined these ideas \cite{gustafsson_physical_2012,gustafsson_physical_2015,jonsson_stored_2015} exploiting the expressions for the reactive stored energy derived by Vandenbosch \cite{vandenbosch_reactive_2010,vandenbosch_simple_2011} and Geyi \cite{geyi_foster_2000,geyi_method_2003}, and they also included magnetic-type antennas. Efficient numerical determination of the optimal current density field by expanding it in terms of the characteristic modes was also recently demonstrated by Chalas and coworkers and by Capek and Jelinek 
and coworkers ~\cite{chalas_computation_2016,capek_optimal_2016,jelinek_optimal_2017,capek_minimization_2017}. Capek \textit{et al.} also recently investigated the role of symmetry in the evaluation of fundamental bounds \cite{capek_role_2021}. Yaghjian has recently proven that the Chu lower bound on Q can be overcome by using highly dispersive material to tune the antenna \cite{yaghjian_overcoming_2018}.

In the literature (e.g., \cite{vandenbosch_reactive_2010,vandenbosch_simple_2011,jonsson_stored_2015}), the antennas are divided into two categories, depending on the features of the current density field they support. Antennas of the {\it electric type} support currents with zero curl, that is, {\it longitudinal} current density fields, while antennas of the {\it magnetic type} support currents with zero divergence, that is, {\it transverse} current density fields. As we shall see, this distinction naturally applies also to plasmonic and high-permittivity resonators.
Plasmonic resonances \cite{maier_plasmonics_2007,novotny_principles_2006} emerge in scatterers made of dispersive materials with a negative real part of the permittivity (metals). In the small-size limit, the plasmonic resonances can be described within the quasi-electrostatic approximation of Maxwell's equations \cite{fredkin_resonant_2003,mayergoyz_electrostatic_2005,tzarouchis_resonant_2017}: they are supported by quasi-electrostatic current density modes, which are longitudinal vector fields. On the other hand, dielectric resonances \cite{tzarouchis_light_2018,koshelev_dielectric_2021} emerge in scatterers made of materials with a high and positive real part of the permittivity. In the small-size limit, the dielectric resonances can be described by the quasi-magnetostatic approximation of Maxwell's equations \cite{forestiere_magnetoquasistatic_2020}: they are supported by quasi-magnetostatic current density modes, which are transverse vector fields.
Quasi-electrostatic and quasi magnetostatic modes are the natural modes of the small-size scatterers \cite{forestiere_time-domain_2021}.
%In both cases, in the limit of high Q and non-interacting quasi-static current modes, the Q factor of the resonant modes is equal to the inverse of the fractional bandwidth of the corresponding resonance peaks in the power spectrum.

This paper tackles the problem of the lower bound of the Q factor for small-size plasmonic and dielectric resonators with arbitrary shape, expanding the current density field induced in the resonator in terms of its quasi-electrostatic or quasi-magnetostatic density resonant modes. This expansion leads to i) the analytical and closed form expressions of the electric and magnetic polarizability tensors of the resonator, whose eigenvalues have been linked to the minimum Q \cite{gustafsson_physical_2007,gustafsson_physical_2012}; ii) the analytical expression of the minimum Q from the dipole moments of the quasistatic current modes of the resonator; iii) the close-form expression of the optimal current.
In particular, the determination of the optimal current
without the use of an optimization procedure or the numerical solution of integral equations, is a considerable advantage over other expansions, such as those based on the characteristic modes \cite{garbacz_modal_1965} of the resonator. This work also unveils the connection between the resonator's minimum Q factor and the Q factor of its natural modes. In particular, we have also found that the minimum Q factor of a resonator with two orthogonal reflection symmetries can be obtained from the Q factors of the single current modes with non-vanishing dipole moments along the major axis through their parallel combination. Moreover, when a plasmonic resonator supports a spatially uniform quasi-electrostatic current mode, this mode is guaranteed to have  the minimum  Q  factor. Due to duality, when a dielectric resonator supports a curl-type quasi-magnetostatic current mode of the form $\QMSuni$ where $\mathbf{c}$ is a constant vector and $\hat{\mathbf{r}}$ is the radial direction, this mode exhibits the minimum Q factor.

The manuscript is organized as follows: in Sect. \ref{sec:Summary} we summarize the definition and main properties of the quasi-electrostatic and quasi-magnetostatic current modes of a small-sized scatterer of arbitrary shape. Then, in Sect. \ref{sec:Link} we address the problem of finding the minimum Q and the corresponding optimal current distribution by expanding the current density field induced in the resonator through its quasistatic current modes. In Sect. \ref{sec:Results} many examples are shown, exemplifying the application of the introduced method to small-sized plasmonic and high-permittivity dielectric resonators of arbitrary shape. In Appendix \ref{sec:AppQ}, we derive the expression of the Q factor for plasmonic and dielectric resonators from their stored energy and radiated power.

\section{Resonances of small-size Scatterers}
\label{sec:Summary}
\begin{figure}
    \centering
    \includegraphics[width=0.4\columnwidth,trim=0 0 0 5]{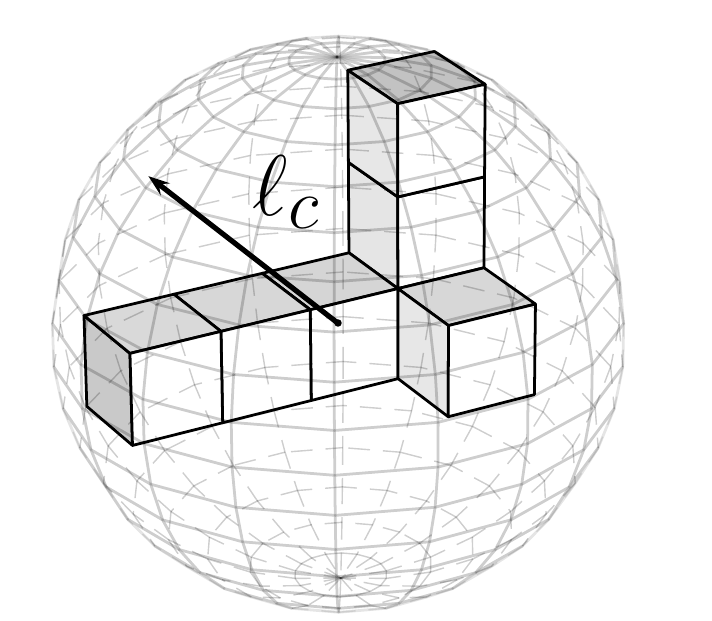}
    \caption{Arbitrarily shaped plasmonic/dielectric resonator enclosed by the circumscribing ``radiansphere" of radius $\ell_c$. In this work we determine the lower bounds of Q factors of small-
size plasmonic and dielectric resonators characterized by quasi-electrostatic and quasi-magnetostatic current modes.}
    \label{fig:lc}
\end{figure}
We consider a linear, homogeneous, isotropic and nonmagnetic scatterer, occupying a volume $V$ with boundary $S=\partial V$ surrounded by vacuum. %, which occupies the external volume $V_e$. 
We define the characteristic linear dimension $\ell_c$ of the scatterer to be the radius of the smallest sphere that surrounds it (Fig. \ref{fig:lc}). We indicate with $\chi(\omega)$ the susceptibility of the scatterer in the frequency domain, which we generally assume to be frequency dispersive. If $\ell_c$ is much smaller than the operating wavelength, resonant electromagnetic scattering can occur due to different mechanisms (e.g., \cite{forestiere_magnetoquasistatic_2020,forestiere_resonance_2020}), including plasmonic and dielectric resonances. %Plasmonic resonances arise from the interplay between the energy stored in the electric field and the polarization energy of matter. Dielectric resonances arise from the interplay between the energy stored in the magnetic field and the polarization energy. %Another important resonant mechanism arises from the interplay between the energy stored in the electric field and the energy stored in the magnetic field (e.g., PEC split ring resonators). 
In the following, we summarize the properties of the resonances and the resonant modes of solid scatterers. In Appendix \ref{sec:SurfRes}, we do the same for surface scatterers (i.e., shells).

\subsection{Plasmonic resonances}

Plasmonic resonances arise in small-size scatterers with a negative real part of the permittivity (e.g., metals). These resonances are associated with the eigenvalues of the integral operator that gives the electrostatic field as a function of the surface charge density on the surface $S$ \cite{fredkin_resonant_2003,mayergoyz_electrostatic_2005},
\begin{equation}
     {\bf j}_h^\parallel \rp  = \gammalo{h}\, \tnabla_\mathbf{r} \oint_{\ptOmega}     \frac{ \n \rpp \cdot  {\bf j}_h^\parallel \left( {\bf r}' \right)}{ 4 \pi \left| \rbt - \rbt' \right| }
       \dS' \qquad \text{in} \ V.
       \label{eq:3Dplasmons}
\end{equation}
 Here, $\Jlo_h\rp$ is a quasi-electrostatic current mode of the scatterer and $\gammalo{h}$ is the corresponding eigenvalue. The eigenvalues of this operator are discrete, real, positive, and size-independent, and $\gammalo{h}\ge 2$ \cite{mayergoyz_electrostatic_2005}. The quasi-electrostatic current modes are {\it longitudinal} vector fields in $V$: they are curl-free and div-free within $V$, but have a non-vanishing normal component on $S$ \cite{mayergoyz_electrostatic_2005}.
%
%\begin{equation}
%    \mathbb{L}_{\|}^{2}(\tOmega)=\left\{\mathbf{w} \in \mathbb{L}^{2}(\tOmega) \mid \nabla \cdot \mathbf{w}^{\|}=0, \nabla \times \mathbf{w}^{\|}=0 \text { in } \tOmega \backslash \partial \tOmega\right\}.
%\end{equation}
%$\mathbb{L}^2$ is the space of square integrable function in $V$%
These current modes are orthogonal, i.e., 
\begin{equation}
  \langle \Jlo_h, \Jlo_k \rangle_{\tOmega} =  \| \mathbf{j}^\parallel_h \|^2 \delta_{h,k}, \quad \forall h,k
  \label{eq:QESorthonorm}
\end{equation}
according to the scalar product
\begin{equation}
   \langle  \mathbf{f} , \mathbf{g}  \rangle_{\tOmega} = \int_{\tOmega}  \mathbf{f}^*   \cdot  \mathbf{g}\,   \dV.
 \label{eq:ScalarProd}
\end{equation}
Moreover, $\mathbf{j}^\parallel_h \left( \mathbf{r} \right)$ satisfies the {\it charge-neutrality} condition $ \oint_{S}    \sigma_h \left( \mathbf{r} \right)  \, \dS = 0$ ,
where
$
    \sigma_h \left( \mathbf{r} \right) =  (i \omega)^{-1} \, \mathbf{j}_h^\parallel \left( \mathbf{r} \right) \cdot \hat{\mathbf{n}}  \left( \mathbf{r} \right)
$
is the surface charge density on $S$ associated with the mode. The electric dipole moment of $\mathbf{j}^\parallel_h \left( \mathbf{r} \right)$ is 
\begin{equation}
  \mathbf{P}_h	 =  \oint_{\Surf}  \sigma_n \rp \mathbf{r} \, \dS =  \frac{1}{i \omega} \int_{V} {\Jlo_{h} \left( \mathbf{r} \right)} \, \dV. 
    \label{eq:Dipole}
\end{equation}

If the mode $\Jlo_h$ has a vanishing electric dipole moment, it is classified as {\it dark}, {\it bright} otherwise \cite{gomez_dark_2013}. If the shape of the resonator has two orthogonal reflection symmetries, the dipole moment of each mode is aligned along one of these directions. When the scatterer has a quasi-electrostatic current mode $\Jlo_h={\bf c}$, that is spatially uniform in $V$ with direction $\bf c$ (as it happens, for instance, in spheres and ellipsoids), the orthogonality condition \ref{eq:QESorthonorm} implies that all the remaining current modes, i.e., $\Jlo_k$ $\forall k \ne h$, have a vanishing electric dipole moment along $\bf c$:
\begin{equation}\label{eq:P_uniformMode}
   {\bf c}\cdot \int_{V} \Jlo_k \rp \dV = \left( i \omega \right) {\bf c} \cdot \PE{k}=0 \qquad \forall k \ne h.
\end{equation}

The current density field $\mathbf{J} \left( \mathbf{r} \right)$ induced in the scatterer by an incident electric field $\mathbf{E}_{inc}$ is given by \cite{mayergoyz_electrostatic_2005,forestiere_resonance_2020}
\begin{equation}
    \mathbf{J} \left( \mathbf{r} \right) \approx  i \omega \varepsilon_0 \sum_{h}   \frac{  \gammalo{h} \, \chi \left( \omega \right)}{ \gammalo{h} + \chi \left( \omega \right) } \langle   \mathbf{j}_h^\parallel, {\bf E}_{inc} \rangle_{\tOmega} \, \frac{\mathbf{j}_h^\parallel \rp}{\| \mathbf{j}_h^\parallel \|^2},
    \label{eq:EQSexpansion}
\end{equation}
where $\varepsilon_0$ is the vacuum permittivity.
The resonance frequency $\omega_h$ of the quasi-electrostatic current mode $\mathbf{j}_h^\parallel$ is the frequency at which the real part of the denominator of Eq. \ref{eq:EQSexpansion} vanishes, i.e.,  \cite{mayergoyz_electrostatic_2005}
\begin{equation}
    \text{Re} \left\{ \chi \left( \omega_h \right) \right\} = -\gammalo{h}.
    \label{eq:ResonanceEQS}
\end{equation}
We now introduce the resonance size parameter $\sizep_h$, defined as
\begin{equation}
    \sizep_h =  \frac{\omega_h}{c_0} \ell_c
    \label{eq:SizeParameter},
\end{equation}
where $c_0$ denotes the light velocity in vacuum. Assuming that the $h$-th current mode $\mathbf{j}_h^\parallel$ is bright and ``isolated" (namely, its resonance frequency is sufficiently far from the resonance frequencies of the other modes), and the dispersion relation of the scatterer is of Drude type, we show in Appendix \ref{sec:AppEQS} that its Q factor has the following expression:
\begin{equation}
  \label{eq:QmodeEQS}
  Q_h^\parallel = 
  -3 \ell_c^3 \, \frac{  \displaystyle\oint_{ \Surf} \sigma_h^* \rp  \cdot \oint_{ \Surf}\frac{ \sigma_h \rpp }{\left| \rt - \rt' \right|  } \dS \dS'} {  \displaystyle\oint_{\Surf} \sigma_h^* \rp \oint_{\Surf}  \sigma_h \rpp \left| \rt - \rt' \right|^2 \dS \dS' } \, \frac{1}{\sizep_h^3},
\end{equation}
which can also be rewritten as
\begin{equation}
Q_h =\frac{1}{\gammalo{h}}  \frac{6 \pi \| \mathbf{j}_h^\parallel \|^2}{\omega_h^2\left\| \mathbf{P}_{h}\right\|^2} \frac{\ell_c^3}{\sizep_h^3}= \frac{1}{\gammalo{h}}  \frac{6 \pi \| \mathbf{j}_h^\parallel \|^2}{\left\| \int_V  \mathbf{j}_{h} \rp \dV \right\|^2} \frac{\ell_c^3}{\sizep_h^3}.
\label{eq:QfactorEQS}
\end{equation}
If the mode is dark, the Q factor presents a more complicated expression which diverges faster than $1/\sizep_h^3$ \cite{forestiere_resonance_2020}. 
\subsubsection{Modal expansion of the electric polarizability tensor}

Following Ref. \cite{jonsson_stored_2016}, the electric polarizability tensor of the scatterer is the linear correspondence,
$
    \TensorE: \left( E_0 \hat{\mathbf{e}} \right) \rightarrow \mathbf{P},
$
between the electric field $\left(E_0 \hat{\mathbf{e}}\right)$ and the electric dipole moment \cite{jackson_classical_1999,yaghjian_force_2018}:
\begin{equation}
  \mathbf{P} = \oint_{\ptOmega} \sigma \rp  \rbt \, \dS,
\end{equation}
where $\sigma\rp$ is the solution of the surface integral equation
\begin{equation}
  \oint_{\ptOmega} \frac{\sigma \rpp}{4 \pi\left|\boldsymbol{r} -\boldsymbol{r}' \right|} \dS'= \left( \varepsilon_0 E_{0}  \hat{\boldsymbol{e}} \right) \cdot \boldsymbol{r} \qquad  \text{on} \,  \ptOmega,
  \label{eq:PolarizabilityE3D}
\end{equation}
subjected to the charge neutrality condition; $E_0$ is a real number and $\hat{\mathbf{e}}$ is a unit vector. %The surface integral operator in Eq.~\ref{eq:PolarizabilityE3D} is closely related to the operator that defines the quasi-electrostatic current modes (Eq. \ref{eq:3Dplasmons}). %One of the main contributions of this paper is the calculation of the polarizability tensor $\TensorE$ of small-sized plasmonic resonators by using the quasi-electrostatic current modes of the object.

We solve Eq. \ref{eq:PolarizabilityE3D} by expanding the unknown $\sigma$ in terms of the surface charge density modes $\sigma_h$, i.e., $ \sigma \rp = \sum_{h} \alpha_h \, \sigma_h \rp,\, \text{on} \,  \ptOmega$.  Substituting this expression in Eq. \ref{eq:PolarizabilityE3D}, which naturally satisfies the charge neutrality condition, multiplying both members by $\sigma_k$, integrating over the surface $S$, and exploiting the orthogonality condition \ref{eq:QESorthonorm}, we obtain the expression for the expansion coefficient $\alpha_k$. Thus, the dipole moment associated with the surface charge density $\sigma$ is given by
\begin{equation}
 \text{\bf P} =   \sum_{h} \frac{\gammalo{h}}{\| \mathbf{j}_h^\parallel \|^2} \PE{h}\otimes \PE{h} \left( \varepsilon_0 E_{0} \right) \hat{\boldsymbol{e}},
  \label{eq:betaPolarizabilityE3D_dip}
\end{equation}
where $\otimes$ denotes the tensor product.  From this, the electric polarizability tensor $\TensorE $ is given by
\begin{equation}
    \TensorE =  \sum_{h}  \frac{\gammalo{h}}{\| \mathbf{j}_h^\parallel \|^2} \, \PE{h} \otimes \PE{h}.
    \label{eq:QESlink}
\end{equation}
This expression is the first important result of this work, as it relates in closed-form the polarizability tensor to the quasi-electrostatic current modes of a plasmonic resonator.

\subsection{Dielectric resonances}
\label{sec:DielectricResonances}
Dielectric resonances arise in small-sized dielectric scatterers with large real part of permittivity. These resonances are associated with the eigenvalues of the magnetostatic integral operator that gives the vector potential as a function of the current density \cite{forestiere_magnetoquasistatic_2020,forestiere_resonance_2020},
\begin{equation}
\Jto_h \rp = \frac{\chi_{h}^\perp }{\ell_c^2}\int_{\tOmega} \frac{ \Jto_h \rpp}{4\pi \left| \rbt - \rbt' \right|}  \dV' \qquad \text{in} \ V,
\label{eq:QMSproblem3D}
\end{equation}
with the condition $ \left. \Jto_h   \cdot \n    \right|_{S} = 0$.
$\Jto_h\rp$ is a quasi-magnetostatic current mode of the scatterer, and $\gammato{h}$ is the corresponding eigenvalue. The above equation holds in {\it weak} form in the functional space of the {\it transverse} vector fields equipped with the inner product \ref{eq:ScalarProd}.
%\begin{multline}\label{eq:L2space}
%    \mathbb{L}^2_\perp \left( \tOmega \right)= %\left\{\mathbf{w} \in \mathbb{L}^{2}(\tOmega) \mid \nabla %\cdot \mathbf{w}^{\perp}=0, \right. \\ \left. \text { in } %\tOmega \backslash \partial \tOmega \text { and } %\mathbf{w}^{\perp} \cdot \hat{\mathbf{n}}=0 \text { on } %\partial \tOmega\right\}.
%\end{multline}
%
The spectrum of the magnetostatic integral operator \ref{eq:QMSproblem3D} is discrete, and the eigenvalues are real and positive \cite{forestiere_magnetoquasistatic_2020}. The quasi-magnetostatic current modes are transverse vector fields defined in $V$: they are div-free in $V$ and have zero normal component on $S$. These modes are orthogonal, i.e.,
\begin{equation}
   \langle \Jto_h , \Jto_k \rangle =  \| \mathbf{j}^\perp_h \|^2 \delta_{h,k},\quad \forall h,k.
    \label{eq:QMSorthonorm}
\end{equation}
The electric dipole moments of the quasi-magnetostatic current modes are equal to zero. The magnetic dipole moment $\PM{h}$ of the mode $\Jto_{h}$ is
\begin{equation}
    \PM{h} = \frac{1}{2} \int_{\tOmega} \rbt \times \Jto_{h} \rp \,  \dV.
    \label{eq:DipoleM}
\end{equation}

If the shape of the scatterer has two orthogonal reflection symmetries, the magnetic dipole moment of each mode is aligned along either one of these directions. If the scatterer supports a mode of the form $\Jto_h  =  \hat{\mathbf{r}} \times \mathbf{c}$, where $\mathbf{c}$ is a constant vector, then the orthogonality condition \ref{eq:QMSorthonorm} implies that all the remaining modes $\Jto_k$ with $ k \ne h$ have a vanishing magnetic dipole moment along $\mathbf{c}$:
\begin{equation}\label{eq:M_uniformMode}
    \int_V \Jto_k \cdot \left( \hat{\mathbf{r}} \times \mathbf{c} \right) \dV = -2\mathbf{c} \cdot \PM{k} = 0, \quad k \ne h.
\end{equation}
For a small dielectric scatterer with high permittivity, the current density field $\mathbf{J} \left( \mathbf{r} \right)$ induced in $V$ by an incident electric field $\mathbf{E}_{inc}$ is \cite{forestiere_magnetoquasistatic_2020}:
\begin{equation}
   {\bf J} \rp \approx  i  \omega \varepsilon_0 \sum_{h}  \frac{ \chi \left( \omega \right) \gammato{h} }{ \gammato{h} - \frac{\omega^2\ell_c^2}{c^2_0}\chi \left( \omega \right)  } \langle  \mathbf{j}^\perp_h, {\bf E}_{inc} \rangle_{\tOmega} \; \frac{\mathbf{j}^\perp_h \left( \rbt \right)}{\| \mathbf{j}^\perp_h \|^2}.
\label{eq:MIMexpansion}
\end{equation}
The resonance frequency of the  quasi-magnetostatic current mode $\Jto_h$ is the frequency $\omega_h$ at which the real part of the denominator of \ref{eq:MIMexpansion} vanishes
\begin{equation}
    \text{Re} \left\{ \chi \left( \omega_h \right) \right\} =\frac{\omega_h^2\ell_c^2}{c^2_0} \gammato{h}.
\end{equation}
In the Appendix \ref{sec:AppMQS} we show that if the $h$-th mode has a non-vanishing magnetic dipole moment, its Q factor is:
\begin{multline}
    Q_h^\perp =  6 \pi \frac{\left\| \mathbf{j}_h^\perp \right\|^2}{\left\|\mathbf{M}_h^\perp\right\|^2} \frac{\ell_c^5}{\gammato{h}} \frac{1}{\sizep_h^3}  \\ = 6  \ell_c^3 \frac{\displaystyle \int_{V} \Jto_h  \rp \cdot \int_{V}  \frac{\Jto_h \rpp}{\left| {\bf r} - {\bf r}' \right|} \dV' \, \dV}{\displaystyle\int_{V} \Jto_h \rp \cdot \int_{V} \Jto_h \rpp \left| {\bf r} - {\bf r}' \right|^2 \dV' \, \dV} \frac{1}{\sizep_h^3}.
    \label{eq:QfactorDiel}
\end{multline}

\subsubsection{Modal expansion of the polarizability tensor}

Following Ref. \cite{jonsson_stored_2016},  the magnetic polarizability tensor $\TensorM$ is the linear correspondence, $ \TensorM: \left( H_0 \hat{\mathbf{e}} \right) \rightarrow \mathbf{M}  $,
between $H_0 \hat{\mathbf{e}}$ ($H_0$ is a real number and  $\hat{\mathbf{e}}$ is a unit vector) and the magnetic dipole moment $\mathbf{M}$ of the current density field $\jlopt$ with zero-average over $\tOmega$ that is solution of the integral equation \cite{jonsson_stored_2015}:
\begin{equation}
 \int_{\tOmega} \frac{\jlopt  \rpp }{4 \pi\left|\rbt - \rbt' \right|}   \dV' = \frac{1}{2} \left( H_{0} \hat{\mathbf{e}}\right) \times \rbt, \qquad \text{in} \ V.
\label{eq:PolarizabilityM3D}
\end{equation}
%The volume integral operator in Eq. \ref{eq:PolarizabilityM3D} is the same operator that occurs in Eq. \ref{eq:QMSproblem3D}, which defines quasi-magnetostatic current modes. 

To solve Eq. \ref{eq:PolarizabilityM3D}, we expand the current density $\jlopt$ in terms of the quasi-magnetostatic current modes. As we have done in the solution of the integral equation \ref{eq:PolarizabilityE3D}, we obtain the expression for $\TensorM$
\begin{equation}
    \TensorM =  \sum_{h}   \frac{\gammato{h}}{\| \mathbf{j}_h^\perp \|^2}  \, \PM{h} \otimes \PM{h}.
    \label{eq:QMSlink}
\end{equation}
As a second important result of this work, this relation expresses in closed form the polarizability tensor as a function of the quasi-magnetostatic current modes.  

\section{Minimum Q factor and Optimal Current Distribution for Plasmonic/Dielectric Resonators}
\label{sec:Link}
We now tackle the problem of determining the optimal current distribution that supports the minimum Q factor for small-sized plasmonic and high-permittivity dielectric resonators.

\subsection{Plasmonic resonators}
\label{sec:Qelectric}

%The Q factor of the single quasi-electrostatic current mode of a small-sized plasmonic resonator is given by expression \ref{eq:QmodeEQS}. 
The problem of finding the minimum Q consists of determining the optimal current density $\jlopt $ in the functional space of longitudinal vector fields defined in $V$, which gives the minimum value of the functional 
\begin{equation}
  \label{eq:QEQS}
  \sizep^3Q = 
  -3 \ell_c^3 \, \frac{  \displaystyle\oint_{ \Surf} \sigma^* \rp  \cdot \oint_{ \Surf}\frac{ \sigma \rpp }{\left| \rt - \rt' \right|  } \dS \dS'} {  \displaystyle\oint_{\Surf} \sigma^* \rp \oint_{\Surf}  \sigma \rpp \left| \rt - \rt' \right|^2 \dS \dS' },
\end{equation}
where $\sizep =  \omega \ell_c/c_0 $ and $\mathbf{j}\cdot\mathbf{n}=\sigma/(i\omega)$. This expression of the Q factor and the following derivation also hold for surface scatterers provided that the quantity $\mathbf{j}\cdot\mathbf{n}$ is replaced by $\tnabla_s \cdot \jlopt $.

Vandenbosch showed that the minimization of functional \ref{eq:QEQS} can be successfully achieved by recasting minimization as the problem of finding the zeros of a matrix determinant \cite{vandenbosch_simple_2011}. Here, we choose to follow the approach of Gustafsson \textit{et al.} \cite{gustafsson_physical_2007,gustafsson_physical_2012,jonsson_stored_2015}.

%They recast the minimization  as the problem of finding the longitudinal current distribution  that yields the minimum electrostatic energy stored in the whole space with a constraint on the square amplitude of the electric dipole moment.  They solve the constrained optimization problem using the Lagrange multiplier method \cite{gustafsson_physical_2007,gustafsson_physical_2012,jonsson_stored_2015}. 

They found that any pair $ \left( \sigma_{opt},\gamma \right)$ satisfying the integral equation \cite{gustafsson_physical_2007,gustafsson_physical_2012,jonsson_stored_2015}
\begin{equation}
     \oint_{\ptOmega} \frac{\sigma_{opt} \rpp}{ 4 \pi \left| \rt -\rt'\right|} \dS'  - \gamma \, \frac{1}{\ell_c^3} \mathbf{r} \cdot  \oint_{\ptOmega} \sigma_{opt} \rpp \rt'  \dS' = 0, \quad \text{on} \ S
     \label{eq:Critical3dE}
\end{equation}
gives a local minimum of the functional $\sizep^3 Q $ ($\gamma$ is the Lagrange multiplier), and among them the absolute minimum can be found. In particular, they found that the minimum of the Q factor is given by \cite{gustafsson_physical_2007,jonsson_stored_2015}:
\begin{equation}
    \left( \sizep^3 Q \right)_\text{min} =  \frac{6 \pi\ell_c^3}{\gamma_\text{e,max}},
    \label{eq:QminQES}
\end{equation}
where $\gamma_\text{e,max}$ is the maximum among the 3 eigenvalues of the electric polarizability tensor $\TensorE$.  Eq. \ref{eq:QminQES} is also consistent with the formula found by Yaghjian and co-workers in Ref. \cite{yaghjian_minimum_2013} (Eq. 44). The corresponding eigenvector returns the direction of the dipole moment $\diropt$ of the optimal surface charge density $\sigma_{opt}$,
\begin{equation}
    \diropt = \int_S \sigma_{opt} \rp \mathbf{r} \, \dS.
    \label{eq:DirOptDef}
\end{equation}

We now expand the optimal current distribution $\jlopt_\text{opt} $  in terms of the quasi-electrostatic current modes $\mathbf{j}^\parallel_h$, $ \jlopt_\text{opt}  \rp = \sum_{h} \alpha_h \, \mathbf{j}^\parallel_h \rp$. We determine the coefficients $\alpha_h$ by substituting this expansion in the critical Eq. \ref{eq:Critical3dE}, using Eq. \ref{eq:DirOptDef} and the orthogonality \ref{eq:QESorthonorm}.
Eventually, we obtain
\begin{equation}
  \jlopt_\text{opt}  \rp = \sum_{h} \frac{\gammalo{h}}{\| \mathbf{j}_h^\parallel \|^2} \left( \diropt \cdot \PE{h}  \right) \mathbf{j}^\parallel_h \rp.
  \label{eq:OptimalCurrent}
\end{equation}
This is a third remarkable result of this work: once the direction of the dipole moment associated with the optimal current is determined, the optimal current is known in closed form. This property constitutes a significant advantage over the previously developed techniques for electrically small antennas, as in Ref. \cite{gustafsson_physical_2007,gustafsson_physical_2012} and Ref. \cite{vandenbosch_simple_2011}, for which the determination of the optimal current is not straightforward because it requires the solution of an integral equation. As we shall see in Sect. \ref{sec:Results}, only a few quasi-electrostatic modes are needed to achieve a good estimate of the optimal current. 

If the shape of the resonator has two orthogonal reflection symmetry planes with normals $\hat{\mathbf{e}}_1$ and $\hat{\mathbf{e}}_2$, the principal axes of $\TensorE$ are the triplet $\left( \ca_1, \ca_2, \ca_3 \right)$, where $\hat{\mathbf{e}}_3$ is orthogonal to both $\hat{\mathbf{e}}_1$ and $\hat{\mathbf{e}}_2$. The dipole moments of the quasi-electrostatic current modes are also aligned along these directions. In this case, the three occurrences of $\TensorE$ are obtained from Eq. \ref{eq:QESlink}. They are given by
\begin{equation}
     {\gamma}_{e,i} = \sum_h \frac{\gammalo{h}}{\| \mathbf{j}^\parallel_h \|^2} \, \left| \hat{\mathbf{e}}_i \cdot \PE{h} \right|^2 
     \qquad i = 1,2,3.
    \label{eq:QESlink2}
\end{equation}
Only the quasi-electrostatic current modes with dipole moment directed along $\ca_i$ contribute to the sum. In this case, by combining Eqs. \ref{eq:QESlink2}, \ref{eq:QminQES}, and \ref{eq:QfactorEQS} we obtain a fourth important result of this work: the minimum Q along the axis $\ca_i$ is given by the parallel formula:
\begin{equation}
    \frac{1}{ \left( \sizep^3 Q \right)_{\text{min},i} } =  \sum_{h_i} \, \frac{1}{ \sizep_{h_i}^3 Q_{h_i}^\parallel }
    \label{eq:QparallelQES3D}
\end{equation}
where the label $h_i$ denotes the modes with non-vanishing electric dipole moments along $\ca_i$. 
As shown in \ref{eq:P_uniformMode}, due to the modes' orthogonality, if there  exists  a  current  mode spatially uniform along $\ca_i$, it is also the only current mode with nonvanishing dipole moment along $\ca_i$, and then it necessarily exhibits the minimum Q factor.
% Due to the orthogonality condition \ref{eq:QESorthonorm}, when  there  exists  a  current  mode  which  is  spatially uniform along $\ca_i$, we can count on the fact that  it is the only current mode with nonvanishing dipole moment along $\ca_i$, and it necessarily exhibits the minimum Q factor.

\begin{figure}[htpb]
\centering
\includegraphics[width=\columnwidth,trim=0 2 0 2]{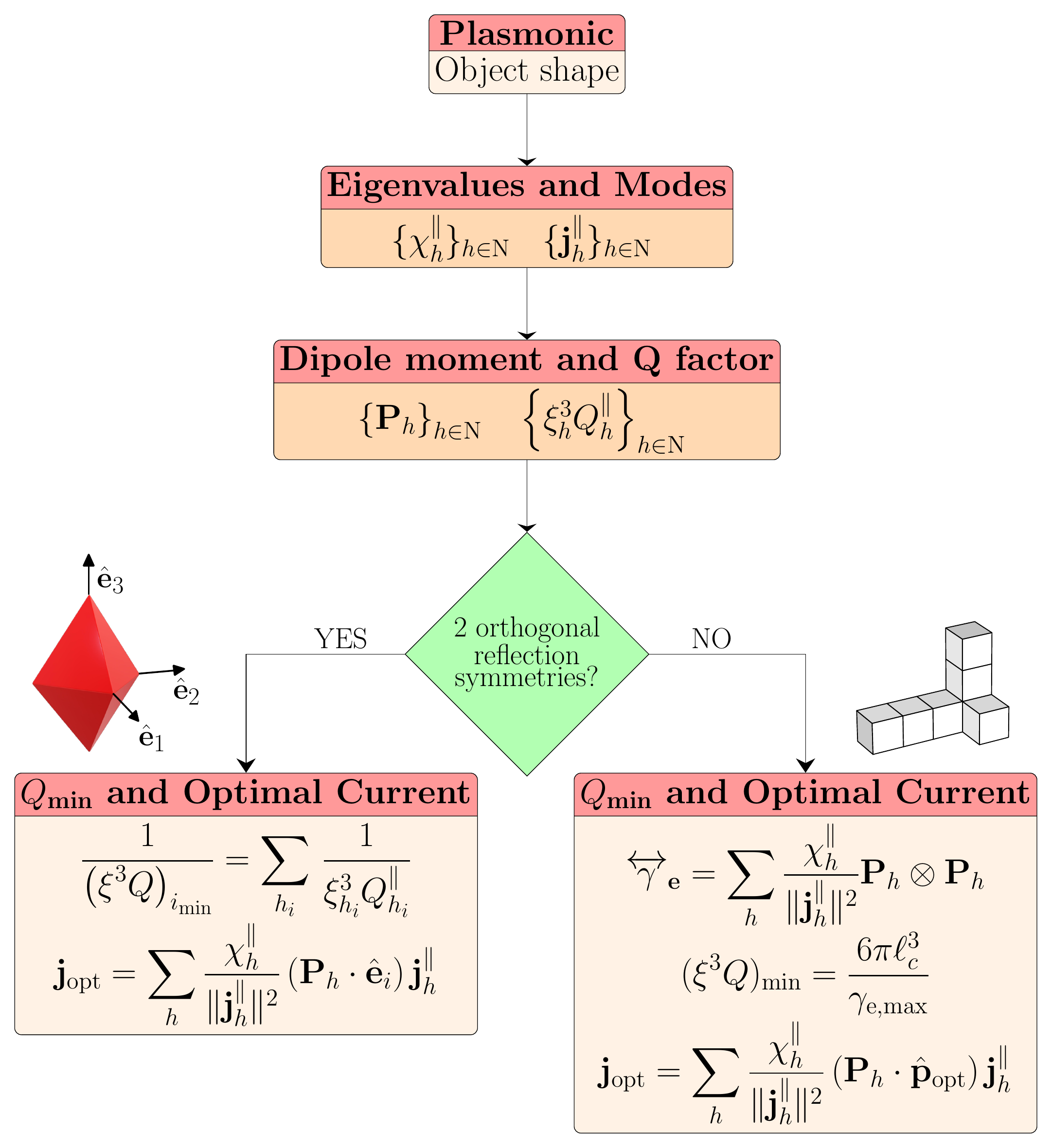} 
\caption{Flow chart for the calculation of the minimum Q factor of an arbitrarily shaped plasmonic resonator using the quasi-electrostatic current modes.  First, we preliminarily calculate the current modes. Then, if the scatterer has two reflection symmetries, the minimum Q along the principal axis of the electric polarizability tensor $\hat{\mathbf{e}}_1$,  $\hat{\mathbf{e}}_2$, $\hat{\mathbf{e}}_3$ is obtained from the Q factor of the current modes with non-vanishing dipole moments oriented along that axis. The absolute minimum Q is the minimum value among these tree values, which corresponds to the optimal current. If no such symmetries are present, then we analytically assembly the polarizability tensor by using the dipole moments and the eigenvalues of the modes, and eventually we find its eigenvalues and eigenvectors. The minimum Q and optimal currents are then immediately obtained. A similar flow chart can be drawn for a high-permittivity dielectric resonator.}
\label{fig:flowchart}
\end{figure}

In conclusion, we summarize in Fig. \ref{fig:flowchart} the algorithm to determine the minimum Q factor and the corresponding optimal current of an arbitrary shaped plasmonic resonator. %First, we  calculate the quasi-electrostatic (plasmonic) current modes associated with the assigned shape on which the search for the optimal current is performed. Then, if the shape has two reflection symmetries, the minimum Q along any of the three principal axis of the electric polarizability tensor is immediately obtained from the Q factor of the single quasi-electrostatic with electric dipole moments oriented along the principal axis. If no such symmetries are present, then we analytically assembly the polarizability tensor by using the electric dipole moments and the eigenvalues of the quasi-electrostatic current modes, and eventually we find its eigenvalues and eigenvectors. The minimum Q and the corresponding optimal current distribution are then straightforwardly obtained.

\subsection{High-permittivity dielectric resonators}

%We consider now a small-size high-permittivity dielectric resonator. The Q factor of a single quasi-magnetostatic current mode is given by expression \ref{eq:QfactorDiel}.
The problem of finding the minimum Q consists in determining the optimal current density $\jtopt $, in the functional space of transverse vector fields defined in $V$, which gives the minimum value of the functional
\begin{equation}
    \sizep^3Q =  6 \ell_c^3 \frac{ \displaystyle\int_V \jlopt   \rp \cdot \displaystyle\int_V  \frac{\jlopt  \rpp}{\left| {\bf r} - {\bf r}' \right|} \dV \, \dV'}{ \displaystyle\int_{V} \jlopt  \rp \cdot \displaystyle\int_{V} \jlopt  \rpp \left| {\bf r} - {\bf r}' \right|^2 \dV \dV'}
\end{equation}
where $\mathbf{j}$ is the current density field. The above expression holds also for surface scatterers of high-conductivity, provided that the volume integrals are replaced by surface integrals.

The minimum Q factor is obtained from the maximum eigenvalue $\gamma_{m,\text{max}}$ of the magnetic polarizability tensor $\TensorM$ \cite{jonsson_stored_2015},
\begin{equation}
    \left( \sizep^3 Q \right)_\text{min} =  \frac{6 \pi \ell_c^3}{\gamma_{m,\text{max}}}.
    \label{eq:QminQMS}
\end{equation}
Since the magnetic polarizability tensor has the closed-form expression \ref{eq:QMSlink}, the determination of $\left( \sizep^3 Q \right)_\text{min}$ only requires the calculation of the eigenvalues of a $3 \times 3$ matrix. Eq. \ref{eq:QminQMS} is also consistent with the formula found by Yaghjian and co-workers in Ref. \cite{yaghjian_minimum_2013}. The eigenvector corresponding to $\gamma_{m,\text{max}}$ returns the direction $\diroptOrt$ of the dipole moment of the optimal current. Following the same steps we have done for the plasmonic resonator, the optimal current is readily obtained in terms of the quasi-magnetostatic current modes,
\begin{equation}
  \mathbf{j}_\text{opt} \rp =  \sum_{h} \gammato{h} \left( \diroptOrt \cdot \PM{h}  \right) \Jto_h \rp.
  \label{eq:OptimalCurrentQMS}
\end{equation}
As we will see in section \ref{sec:Results}, in many scenarios, only a few current modes have to be considered to have a good estimation of the minimum Q factor. 

As for the plasmonic resonators, if the shape of the resonator has two orthogonal reflection symmetry planes with normals $\ca_1$ and $\ca_2$, the principal axis of $\TensorM$ are the triplet $\left( \ca_1, \ca_2, \ca_3 \right)$, where $\ca_3$ is orthogonal to both $\ca_1$ and $\ca_2$. Thus, the three occurrences of $\TensorM$ are
\begin{equation}
    \gamma_{m,i} =  \sum_h  \frac{\gammato{h}}{\| \mathbf{j}_h^\perp \|^2} \, \left| \hat{\mathbf{e}}_i \cdot \PM{h} \right|^2 \qquad i = 1,2,3.
    \label{eq:QMSlink2}
\end{equation}
where the summation runs only over the quasi-magnetostatic current modes with magnetic dipole moment directed along $\ca_i$. The minimum Q along the axis $\ca_i$ is then obtained by their parallel combination
\begin{equation}
    \frac{1}{ \left( \sizep^3 Q \right)_{\text{min},i} } =  \sum_{h_i}  \frac{1}{ \sizep_{h_i}^3 Q_{h_i}^\perp }
    \label{eq:QparallelQMS3D}
\end{equation}
where only the modes with magnetic dipole moment directed along $\ca_i$ have to be considered. 
In addition, as shown in \ref{eq:M_uniformMode}, due to the modes' orthogonality, if  there  exists a  current  curl-type mode  in the form $\mathbf{r} \times \mathbf{c}$,  it is the  only one  with  non-vanishing magnetic dipole  moment along direction $\mathbf{c}$. Thus, it necessarily has the minimum Q factor.
% In addition, due to the orthogonality of the quasi-magnetostatic current modes, if  there  exists a  current  curl-type mode  of , in the form $\mathbf{r} \times \mathbf{c}$,  this  mode  is the  only one  with  non  vanishing magnetic dipole  moment along direction $\mathbf{c}$. Thus, it necessarily has the minimum Q factor.

\section{Results and Discussion}
\label{sec:Results}

We now exemplify the outlined method, by evaluating the minimum Q factor of small-sized plasmonic and dielectric resonators with different shapes. We first consider shapes that support uniform quasi-electrostatic   modes and curl-type quasi-magnetostatic  modes, which are guaranteed to have the minimum Q factor. Then, we consider shapes with two orthogonal reflection symmetries, where the minimum Q factor can be obtained from the Q factors of the quasi-static current modes through the parallel formula. Eventually, we consider shapes with no symmetry. The electrostatic eigenvalue problem Eqs.~\ref{eq:3Dplasmons} is solved by the numerical method outlined in Refs. \cite{fredkin_resonant_2003,mayergoyz_analysis_2007}, and the magnetostatic eigenvalue problem Eqs.~  \ref{eq:QMSproblem3D} is solved by the numerical method  described in Refs. \cite{forestiere_magnetoquasistatic_2020} and \cite{forestiere_electromagnetic_2019}.

\begin{figure*}[h!]
\centering
\includegraphics[width=\textwidth]{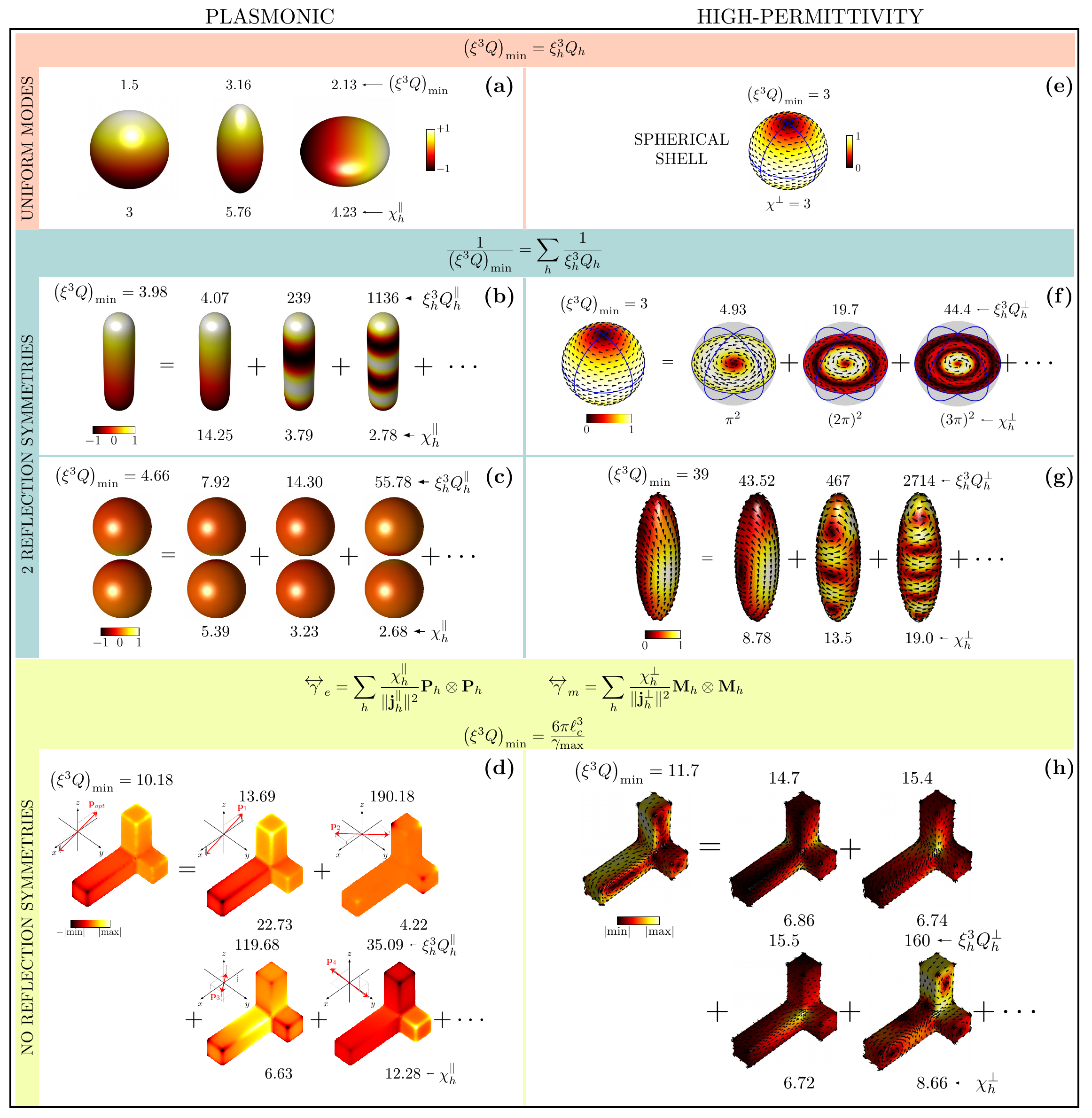} 
\caption{Minimum Q and corresponding optimal charge/current distribution supported by plasmonic $\bf (a-d)$ and dielectric $\bf(e-h)$ resonators. {\bf Plasmonic resonators.} $\bf(a)$ Optimal charge density supported by a sphere and by prolate and oblate spheroids with aspect ratio $2:1$; the bright modes of these shapes are the uniform current modes. Optimal charge density supported by geometries exhibiting two reflection symmetries, namely a rod $\bf(b)$ and a sphere's dimer $\bf(c)$ and, by a shape without symmetries $\bf(d)$. In (b-d), on the right of the ``$=$" sign, plasmonic modes with lowest Q factor, their Q factor (top) and eigenvalues (bottom). The colormap represents the electric charge density. {\bf Dielectric resonators.} $\bf(e)$ Current density mode of a spherical shell of the form $\hat{\mathbf{r}} \times \mathbf{c}$ that is the optimal current density for the spherical shell. Optimal current density supported by geometries exhibiting two reflection symmetries, namely a solid sphere $\bf(f)$ and an ellipsoid $\bf(g)$, and by a shape with no symmetries $\bf(h)$. In each panel, on the right of the ``$=$" sign, quasi-magnetostatic  modes with lowest Q factor, their individual Q factor (top) and eigenvalues (bottom). The colormap represents the magnitude of the current density, the arrows its direction.}
\label{fig:Modes_Table3D}
\end{figure*}

\subsection{Plasmonic resonator}
\label{sec:3Delectric}

{\bf Shapes with uniform current modes -}
\label{sec:UniformE} A sphere with unit radius has three degenerate quasi-electrostatic uniform current modes, one for each of the three orthogonal direction $\hat{\mathbf{x}},\hat{\mathbf{y}},\hat{\mathbf{z}}$, with eigenvalues $\chi_{x,y,z}^\parallel = 3$. We show the surface charge density of 
the current mode $\mathbf{j}^\parallel_z = \sqrt{ 3 / \left( 4\pi \right)} \, \hat{\mathbf{z}}$
in Fig. \ref{fig:Modes_Table3D}(a). For the considerations made in Sect. \ref{sec:Qelectric}, these modes are the {\it only} bright modes supported by a sphere. Thus, their Q factors
coincide with the minimum Q factor supported by the sphere for longitudinal currents,
\begin{equation}
    \left( \xi^3  Q  \right)_\text{min} =  \sizep_{x,y,z}^3 Q^\parallel_{x,y,z} = 1.5\, .
\end{equation}

Similarly, a  rotationally symmetric spheroid (around $\hat{\mathbf{z}}$), has three uniform current modes, $\mathbf{j}^\parallel_x = j_0 \hat{\mathbf{x}}$, $\mathbf{j}^\parallel_y = j_0 \hat{\mathbf{y}}$, and $\mathbf{j}^\parallel_z = j_0 \hat{\mathbf{z}}$, where $j_0=1/\sqrt{V}$ and $V=\frac{4}{3}\pi a_x^2 a_z$, where and $a_x$ and $a_z$ are the semi-axis. They are the {\it only} bright modes of the spheroid. 
The expression \ref{eq:QfactorEQS} for the Q factor in this case simplifies to:
\begin{equation}
     \sizep_{x,y,z}^3 Q^\parallel_{x,y,z} =\frac{6\pi \ell_c^3}{ V} \frac{1}{\chi_{x,y,z}^\parallel}.
\end{equation}
The eigenvalue $\chi^\parallel_h$ corresponding to the current mode aligned along the major axis is the maximum eigenvalue, therefore the minimum Q is associated with it. As an example, in Fig. \ref{fig:Modes_Table3D}(a), we consider the case of a prolate and an oblate spheroid with aspect ratio $2:1$.

{\bf Shapes with non-uniform current modes and two reflection symmetries}. We consider a rod with radius $R$ and height $H=4R$, aligned along $\hat{\mathbf{z}}$. We modeled the rod as a superellipsoid, with boundary $ \left( x/R \right)^{2} + \left( y/R \right)^{2} + \left( z/(4R) \right)^{10} = 1$. We follow the algorithm outlined in Fig. \ref{fig:flowchart}. In Figure \ref{fig:Modes_Table3D} (b), on the right of the ``$=$" sign, we show the surface charge density of the three bright modes with lowest Q and with electric dipole moment directed along $\hat{\bf z}$. We obtain the minimum Q factor by combining the Q factor of the bright modes by using the parallel formula and the optimal current by applying Eq. \ref{eq:OptimalCurrent}. In the same figure, on the left of the ``$=$" sign we show the charge density corresponding to the minimum Q factor.  The Q factor of the first current mode is very close to the Q bound because for the considered superellipsoid, the first current mode on the right of the equal sign is almost uniform. The relative error in the calculation of $\left( \xi^3\,Q \right)_ \text{min}$ by considering only the first three current modes is below $0.2\%$.

In Fig. \ref{fig:Modes_Table3D} (c), we also consider the case of a sphere dimer of radius $R$, aligned along the $\hat{\mathbf{z}}$-axis with an edge-edge gap $\delta=R/10$. Similarly to the rod, the minimum Q factor is obtained by combining the Q factor of the bright modes with electric dipole moments aligned along the $\hat{\mathbf{z}}$ axis, by using the parallel formula. On the other hand, for the sphere dimer the first mode exhibits a Q that is quite larger than the minimum. This is because the dimer of two nearly touching spheres supports modes that strongly deviate from the uniform distribution \cite{pascale_full-wave_2019}. The relative error in the calculation of $\left( \xi^3\,Q \right)_ \text{min}$ by considering only the first three modes is below $0.3\%$.

\textbf{Shapes with non uniform current modes and no symmetries -} We consider a block with three arms of different lengths. We first compute the quasi-electrostatic modes of this scatterer. The four bright modes with lowest Q factor are shown in Fig. \ref{fig:Modes_Table3D} on the right of the ``$=$" sign. The direction of the dipole moment $\PE{h}$ of each mode is also shown in the insets. Considering that there are no symmetries, we have to preliminary assemble the polarizability tensor using Eq. \ref{eq:PolarizabilityE3D} and find its maximum eigenvalue. On the left of Fig. \ref{fig:Modes_Table3D}(d), we show the surface charge density associated with the optimal current obtained by Eq. \ref{eq:OptimalCurrent}.  The relative error in the estimation of $\left( \xi^3\,Q \right)_ \text{min}$ by taking into account only the four modes shown in Fig. \ref{fig:Modes_Table3D}(d) is $26\%$. We have to consider at least 25 modes to have an error below $10\%$.

\subsection{Dielectric Resonator}
\label{sec:3Dmagnetic}
\textbf{Shapes supporting a quasi-magnetostatic curl-type mode -}
We now consider a dielectric resonator having the shape of a spherical shell with unit radius. First, we compute the quasi-magnetostatic current modes associated with this shape by Eq. \ref{eq:QMSproblemSurf} of the Appendix \ref{sec:SurfResMQS}. This shape supports three degenerate current curl-type modes with non-zero magnetic dipole moment:
$
    \Jto_{\hat{\mathbf{c}}} = \sqrt{{3}/(2\pi)} \, \hat{\mathbf{r}} \times \hat{\mathbf{c}}
$
where $\hat{\bf c} = \hat{\bf x}, \hat{\bf y}, \hat{\bf z}$; the magnetic dipole moment is oriented along $\hat{\mathbf{c}}$ and $\chi^\perp_{\hat{\bf c}} = 3$. According to the discussion of Sect. \ref{sec:DielectricResonances}, they are the only modes with non-vanishing magnetic dipole moment. We show one of these current modes in Fig. \ref{fig:Modes_Table3D}(e).  Thus, applying Eq. \ref{eq:QparallelQMS3D} the minimum Q factor is
\begin{equation}
    \left( \sizep^3 Q \right)_\text{min} =     \left( \sizep_{\hat{\mathbf{c}}}^3 Q_{\hat{\mathbf{c}}}^\perp \right) = 3.
\end{equation}
This is in agreement with Refs. \cite{thal_new_2006,vandenbosch_simple_2011,gustafsson_physical_2015}.

\textbf{Shapes with two reflection symmetries -}
We now consider a dielectric sphere resonator of unit radius. Unlike the spherical shell, the solid sphere does not support a mode of the form $\QMSuni$. We compute the supported quasi-magnetostatic current modes solving the eigenvalue problem \ref{eq:QMSproblem3D}. We limit our analysis to the ones having non-vanishing magnetic dipole moment along the $\hat{\mathbf{z}}$ axis:
$ \mathbf{j}_{h} (r, \theta, \phi) =
{\sqrt{3 \pi}}/{2} \, j_1 \left( h \pi r \right) \, \hat{\mathbf{r}} \times \hat{\mathbf{z}}, $ where $h \in \mathbb{N}$ and  $j_1$ is the spherical Bessel function of the first kind and order $1$. They are associated to the eigenvalues
$\gammato{h} =  \left( h \pi \right)^2$. The
 Q factors of the modes are
$\left( \sizep_h^3 Q_h^\perp \right) =  {\left( h \pi \right)^2}/{2}.$
The  first three current modes are shown on the right of the ``$=$" sign in Fig. \ref{fig:Modes_Table3D}(f), with their Q factor. The minimum Q factor $\left( \sizep^3 Q \right)_\text{min}$ is obtained by applying \ref{eq:QparallelQMS3D}:
\begin{equation}
    \left( \sizep^3 Q \right)_\text{min}^{-1} = \sum_{h} \left( \sizep_h Q_h^\perp \right)^{-1} = \frac{2}{\pi^2} \sum_{h} \frac{1}{h^2} =  \frac{2}{\pi^2} \frac{\pi^2}{6} = \frac{1}{3},
\end{equation}
which is in agreement with Thal's analysis.
In this parallel, by only considering the first 4 modes, we obtain an error of $15.4 \%$; we have to consider at least $13$ modes to have an error below $5\%$. The current density field is obtained by applying Eq.~\ref{eq:OptimalCurrentQMS}:
% . It is localized on the sphere's boundary
\begin{equation}
    \mathbf{j}_{opt} (r, \theta, \phi) = \sqrt{\frac{3}{2\pi}}  \, \delta \left( r - 1 \right)\,\hat{\mathbf{r}} \times \hat{\mathbf{z}},
\end{equation}
where $\delta$ is a Dirac delta function. Thus, it corresponds to a surface current localized on the sphere's surface, which is the same optimal current found for a spherical shell. %In conclusion, the minimum Q factor and the corresponding optimal current are in agreement with the results achieved by Thal \cite{thal_new_2006} for a spherical inductor. 

We now consider a spheroidal shell with aspect ratio $4:1$, with major axis aligned along $\hat{\mathbf{z}}$. Also this shape does not support a curl-type mode. We compute the quasi-magnetostatic resonances by solving the eigenvalue problem \ref{eq:QMSproblemSurf} of the Appendix \ref{sec:SurfResMQS}. The minimum Q factors is associated with the set of quasi-magnetostatic modes exhibiting a non-vanishing magnetic dipole moment along the major axis.  In Fig. \ref{fig:Modes_Table3D}(g), we show the optimal surface current, the eigenvalues, and Q factor of the three current modes with the lowest Q factor. The value of the minimum Q factor is obtained using Eq. \ref{eq:QparallelQMS3D}: $\left( \xi^3\,Q \right)_\text{min} = 39$. If we only consider the three modes shown in Fig. \ref{fig:Modes_Table3D}(g), an error $<1 \%$ is obtained.

\textbf{Shapes with no symmetries -} We consider a shell with no reflection symmetries, defined as the boundary of a block with three arms of different lengths. We preliminarily compute its quasi-magnetostatic current modes by solving Eq. \ref{eq:QMSproblemSurf}, their magnetic dipole moments $\PM{h}$ by Eq. \ref{eq:DipoleM}, and Q factors by Eq. \ref{eq:QfactorDiel}. The four modes with the lowest Q factor are shown in Fig. \ref{fig:Modes_Table3D}(h) on the right of the equality sign, with their Q factor (above), and eigenvalue (below). We assembly the magnetic polarizability tensor $\TensorM$ using Eq. \ref{eq:QMSlink} from the dipole moments of the current modes $\PM{h}$. The maximum eigenvalue $\gamma_{max}$ of $\TensorM$ gives the minimum Q factor through Eq. \ref{eq:QminQES}. The optimal current obtained by using Eq. \ref{eq:OptimalCurrentQMS} is shown on the left of the equality sign in Fig. \ref{fig:Modes_Table3D}(h). Only by considering the first 3 modes, we obtain an estimate of $\left(\sizep^3 Q\right)_\text{min}$ with an error of $12\%$.

\section{Conclusions}

We have tackled the problem of finding the minimum Q and the optimal current of electrically small plasmonic and high-index nano-resonators, a topic of great relevance for the growing metamaterials and nano-optics community.
We show that this electromagnetic problem is conveniently described in a basis formed by the quasi-static resonance modes supported by the scatterer, which are the natural modes of the resonator in the small-size limit. %Specifically, quasi-electrostatic  modes are  the natural modes  of  small  scatterers with a negative permittivity (e.g., metals) while  quasi-magnetostatic  modes are the natural modes of small scatterers with high and positive real part of permittivity. 
We demonstrated that the expansion of the current density in terms of quasistatic modes leads to analytical closed form expressions for the electric and magnetic polarizability tensors, whose  eigenvalues  are directly  linked  to  the  minimum  Q.  Hence, we have been able to determine the minimum Q and the corresponding optimal current distributions in the scatterers in closed form. In particular, we found that, when the resonator exhibits two orthogonal reflection symmetries, its minimum Q factor can be simply obtained from the Q factors of the quasistatic modes of the radiator with non-vanishing dipole moment along with the major axis.  Moreover, when a plasmonic resonator supports a spatially uniform quasi-electrostatic current mode, this mode is guaranteed to have  the  minimum  Q  factor.   Because of duality,  when  a  dielectric resonator supports a quasi-magnetostatic current, curl-type mode, in form $\QMSuni$ where $\mathbf{c}$ is a constant vector and $\hat{\mathbf{r}}$ is the radial direction, this mode also exhibits the minimum Q factor. The introduced method can be also applied to find the minimum Q of translational invariant scatterers \cite{pascale_bandwidth_2021}.

 In this manuscript we considered plasmonic and high-permittivity resonant scatterers, limiting the search space for optimal currents either to longitudinal or transverse current density vector fields,
\cite{capek_optimal_2016}. However, in principle, lower bounds may be obtained by simultaneously considering both type of vector fields \cite{capek_optimal_2016}, e.g., in dual mode antennas \cite{gustafsson_physical_2015}.

Beyond the limit of small-size resonators, the advantages of the quasistatic basis become less relevant, and the use of convex optimization over current density becomes necessary \cite{chalas_computation_2016,capek_optimal_2016,jelinek_optimal_2017,capek_minimization_2017} to find the minimum Q. Nevertheless, since the quasi-electrostatic and quasi-magnetostatic modes form a basis for the square-integrable currents defined within the scatterer, they can be used to represent the optimal current solution of convex optimization problems. We expect that the optimal current in Drude plasmonic particle will no longer be irrotational (as in the small-particle limit). Dually, the optimal current in high-index resonator will no longer be solenoidal. In both cases, the contribution of both electrostatic and magnetostatic mode will be needed to determine the optimal current; nevertheless, if the size of the scatterer is smaller or comparable to the resonance frequency, we expect that only few modes will be required to describe the optimal current.

The introduced framework bridges a classic antenna problem to the field of resonant scattering, and in particular to the field of plasmonics, metamaterials and nano-optics. Our results may be especially appealing to researchers and engineers working in photonics and polaritonics, leading to optimal solutions to enable enhanced light-matter interactions through engineered nanostructures.

\appendix

\subsection{Q factor of small scatterers}
\label{sec:AppQ}
In this Appendix, we derive the expression of the Q factor for plasmonic and high-permittivity resonators with characteristic dimension $\ell_c$ much smaller than the operating wavelength $\lambda$, $\xi=2\pi \ell_c/\lambda \ll 1$. The Q factor of a self-resonant structure is defined as $2 \pi$ times the ratio between the mean value over the cycle of the stored energy $\mathscr{W}_\text{stored}$ and the energy $\mathscr{W}_\text{lost}$ lost per cycle by damping processes, both evaluated at the resonance frequency of the h-th mode $\omega_h$ (e.g., \cite{noauthor_ieee_2014,schab_energy_2018}),
\begin{equation}
    \displaystyle Q\ {\stackrel {\mathrm {def} }{=}}\ 2\pi \times {\frac {\mathscr{W}_\text{stored}}{\mathscr{W}_\text{lost}}}= \omega_h \times \frac{\mathscr{W}_\text{stored}}{\text{power loss}}.
    \label{eq:QfacDef}
\end{equation}
%In papers on tuned PEC antennas, such as \cite{vandenbosch_simple_2011,gustafsson_physical_2007}, the quality factor appears to be twice this expression. In these papers, the stored energy is not the total stored energy in the resonance since it does not include the energy stored in the series impedance. 

\subsubsection{Plasmonic resonator}
\label{sec:AppEQS}
In this section, we evaluate the stored energy, the radiated power, and the Q factor of a dispersive plasmonic scatterer. We describe the metal by using the Drude model (e.g., \cite{maier_plasmonics_2007}) with vanishing dissipation losses,
\begin{equation}
    \chi\left( \omega \right) = -\frac{\omega_p^2}{\omega \left( \omega + i \nu \right)} \approx -\frac{\omega_p^2}{\omega^2};
    \label{eq:ChiMetal}
\end{equation}
$\omega_p$ is the plasma frequency and $\nu$ is the damping rate of the free electrons of the metal, which is assumed to be much smaller than $\omega_p$ and the operating frequency $\omega$.
\label{sec:Qplasmon}

{\bf Mean value of the stored energy -}
\label{sec:WstoredE}
The electrostatic field $ \mathbf{E}_h$ associated with the surface charge density $\sigma_h$ of the quasi-electrostatic (plasmon) current mode $\mathbf{j}^\parallel_h$  is given by \cite{mayergoyz_electrostatic_2005}
\begin{equation}
    \mathbf{E}_h \left( \mathbf{r} \right) = -\frac{\nabla_\mathbf{r}}{4\pi\varepsilon_0}\oint_{\Surf} \frac{\sigma_h \left( \mathbf{r}' \right)}{\left| \mathbf{r} - \mathbf{r}'  \right|}  dS'.
    \label{eq:Efield}
\end{equation}
The mean value of the energy stored in the whole space $\mathscr{W}_\text{stored}$ in the presence of the metal particle is dominated by its electric share $\mathscr{W}_\text{stored}^{\left(e\right)}$, which is given by \cite{brillouin_wave_2013,landau_electrodynamics_2013,jackson_classical_1999,yaghjian_impedance_2005}:
\begin{multline}
    \label{eq:Wpol}
    \mathscr{W}_\text{stored}^{\left(e\right)} = \\ \frac{\varepsilon_0 }{4} \left( 1 + \frac{\partial \left( \omega \chi  \right)}{\partial \omega} \right) \int_V \left\| \mathbf{E}_h \right\|^2\dV +  \frac{\varepsilon_0 }{4}  \int_{\Vext} \left\| \mathbf{E}_h \right\|^2 \dV,
\end{multline}
where $\Vext$ is the external space.
Using the following identity (see Eq. 35 in Ref. \cite{mayergoyz_electrostatic_2005,yaghjian_overcoming_2018})
\begin{equation}
    \int_{\Vext} \left\| \mathbf{E}_h \right\|^2\dV  = \left( \gammalo{h}-1\right) \int_{V}  \left\| \mathbf{E}_h \right\|^2\dV .
\end{equation}
 in Eq. \ref{eq:Wpol}, we obtain: 
\begin{equation}
     \mathscr{W}_\text{stored}^{\left(e\right)} =  \frac{\varepsilon_0}{4} \left(\gammalo{h} + \frac{\partial \left( \omega \chi  \right)}{\partial \omega}  \right) \int_{V} \left\| \mathbf{E}_h \right\|^2\dV.
    \label{eq:Wpol2}
\end{equation}
Then, we evaluate the norm of the electric field in $V$
\begin{equation}
    \int_{V} \left\| \mathbf{E}_h \right\|^2\dV  =
    \frac{1}{4\pi \varepsilon_0^2 {\gammalo{h} } } \oint_{\Surf} {\sigma_h^* \left( \mathbf{r} \right) }\oint_{\Surf} \frac{\sigma_h \left( \mathbf{r}' \right)}{\left| \mathbf{r} - \mathbf{r}'  \right|}  dS' \, dS
    \label{eq:Enorm}
\end{equation}
where we have used Eq. \ref{eq:Efield}, the divergence theorem, and  $\displaystyle \mathbf{E}_h \cdot \hat{\mathbf{n}} = -{\sigma_h}/{(\varepsilon_0\gammalo{h})}$ $ \text{on}  \,  \Surf$.

In conclusion, by combining \ref{eq:Enorm} and \ref{eq:Wpol2} we obtain:
\begin{multline}
     \mathscr{W}_\text{stored}^{\left(e\right)} =   \beta \left( \omega \right) \frac{1}{8\pi \varepsilon_0}  \oint_{\Surf} {\sigma_h^* \left( \mathbf{r} \right) }\oint_{\Surf} \frac{\sigma_h \left( \mathbf{r}' \right)}{\left| \mathbf{r} - \mathbf{r}'  \right|}  dS' \, dS %= \\ \beta \left( \omega \right) \mathscr{W}_{e} 
    \label{eq:WstoredFin}
\end{multline}
where
\begin{equation}
\beta \left( \omega \right) = \frac{1}{2 \gammalo{h}} \left( \gammalo{h} +  \frac{\partial \left( \omega \chi  \right)}{\partial \omega}  \right).
\end{equation}

For a Drude metal with vanishing losses (see Eq. \ref{eq:ChiMetal}) we have $     \frac{\partial \left( \omega \chi  \right)}{\partial \omega} = - \chi \left( \omega \right)$.  
% \begin{equation}
%      \frac{\partial \left( \omega \chi  \right)}{\partial \omega} = - \chi \left( \omega \right);
% \end{equation}
Moreover by Eq. \ref{eq:ResonanceEQS}, at the resonance we have $\chi \left( \omega_h \right) \approx -\gammalo{h}$, thus
$ \beta \left( \omega_h \right) \approx 1 $.
In conclusion, we obtain:
\begin{equation}
     \mathscr{W}_\text{stored}^{\left(e\right)} =   \frac{1}{8\pi \varepsilon_0}  \oint_{\Surf} {\sigma_h^* \left( \mathbf{r} \right) }\oint_{\Surf} \frac{\sigma_h \left( \mathbf{r}' \right)}{\left| \mathbf{r} - \mathbf{r}'  \right|}  dS' \, dS.
     \label{eq:WstoredPlas}
\end{equation}
%We note that the stored energy of an electrically small plasmonic radiator is twice the expression of the stored electric energy obtained by Vandenbosch \cite{vandenbosch_reactive_2010} for a non-dispersive PEC electric radiator.

{\bf Radiated power -}
\label{sec:WradE}
The power $\mathscr{P}_h$ radiated in free-space by the quasi-electrostatic current mode $\mathbf{j}^\parallel_h$ with non-zero electric dipole moment is
\begin{multline}
    \mathscr{P}_h = \frac{\mu_0 \omega_h^4}{12 \pi c}  \left\| {\bf P}_h \right\|^2 =  \frac{\omega}{2} \frac{ \left( k_0 \ell_c \right)^3 }{6 \pi \varepsilon_0} \frac{1}{\ell_c^3} \left\| {\bf P}_h \right\|^2 = \\ \frac{\omega_h}{3} \frac{\left( k_0 \ell_c\right)^3}{\ell_c^3}\frac{1}{8 \pi \varepsilon_0 } \oint_{\Surf} \sigma_h^*  \rp \oint_{\Surf} \sigma_h  \rpp \left| \rt - \rt' \right|^2 \dS \dS'
    \label{eq:PradEQS}
\end{multline}
where we used the identity $\displaystyle\left\| {\bf P}_h \right\|^2 = - \frac{1}{2} {  \oint_{S} \sigma_h^* \rp \oint_{S}  \sigma_h \rpp \left| \rt - \rt' \right|^2 \dV'\dV } $.

{\bf Q factor -}
\label{sec:Qe}
We now compute the Q factor by using definition \ref{eq:QfacDef}, assuming negligible dissipation losses in the material. By combining Eqs. \ref{eq:WstoredPlas} and \ref{eq:PradEQS} we get Eq. \ref{eq:QmodeEQS}.
By exploiting the following identities%
\begin{align}
     &\frac{1}{8\pi \varepsilon_0}  \oint_{\Surf} {\sigma_h^* \rp}\oint_{\Surf} \frac{ \sigma_h \rpp }{\left| \mathbf{r} - \mathbf{r}'  \right|}  \dS \,\dS' = 
     \frac{\omega^2}{2 \varepsilon_0} \frac{\| \mathbf{j}^\parallel_h \|^2}{\chi^\parallel_h},&\nonumber \\
    &- \frac{1}{2} {  \int_{V} \sigma_h^* \rp \int_{V}  \sigma_h \rpp \left| \rt - \rt' \right|^2 \dV'\dV } \qquad&\nonumber\\ 
    &\qquad\qquad\qquad\qquad\qquad\quad= \omega^2  \left\| \int_V \mathbf{j}_h \rp \dV   \right\|^2,&
\end{align}
% \begin{equation}
%      \frac{1}{8\pi \varepsilon_0}  \oint_{\Surf} {\sigma_h^* \rp}\oint_{\Surf} \frac{ \sigma_h \rpp }{\left| \mathbf{r} - \mathbf{r}'  \right|}  \dS \,\dS' = 
%      \frac{\omega^2}{2 \varepsilon_0} \frac{\| \mathbf{j}^\parallel_h \|^2}{\chi^\parallel_h},
%      \end{equation}
% \begin{multline}
%     - \frac{1}{2} {  \int_{V} \sigma_h^* \rp \int_{V}  \sigma_h \rpp \left| \rt - \rt' \right|^2 \dV'\dV } \\ = \omega^2  \left\| \int_V \mathbf{j}_h \rp \dV   \right\|^2,
% \end{multline}
Eq. \ref{eq:QmodeEQS} becomes Eq. \ref{eq:QfactorEQS}.
It is worth noting that the expression \ref{eq:QmodeEQS} coincides with the one obtained by Vandenbosch \cite{vandenbosch_simple_2011} for an electrically small non-dispersive tuned PEC radiator of the electric-type.

\subsubsection{High-permittivity dielectric resonator}
\label{sec:AppMQS}
We now evaluate the stored energy, the radiated power, and the Q factor of a high-permittivity dielectric resonator with non-dispersive susceptibility $\chi \left( \omega \right) = \chi_0 \gg 1$ in the frequency range of interest.

{\bf Mean value of the stored energy -}
\label{sec:StoredEQS}
The stored energy $\mathscr{W}_\text{stored}$ is the sum of the stored electric and magnetic energies. Starting from Eq. \ref{eq:Wpol},  the stored electric energy $\mathscr{W}^{\left(e\right)}_\text{stored}$ can be rewritten as
\begin{multline}
    \mathscr{W}_\text{stored}^{\left(e\right)} =  \frac{\varepsilon_0}{4}  \left( 1 + \frac{\partial \omega \chi_0}{\partial \omega}\right) \int_V \left\| \mathbf{E}_h \right\|^2\dV\\ +  \frac{\varepsilon_0}{4}  \int_{\Vext} \left\| \mathbf{E}_h \right\|^2\dV  \approx \frac{\mu_0 \ell_c^2}{4\gammato{h}} \left\|\mathbf{j}_h\right\|^2,
\end{multline}
where we have exploited the fact that the second term dominates over the remaining two for $\chi_0 \uparrow \infty$, and the identity 
\begin{equation}
    \int_{V}  \left\|\mathbf{E}_h\right\|^2 \dV = \ell_c^2 \mu_0 \, \frac{\left\|\mathbf{j}_h\right\|^2}{\gammato{h} \varepsilon_0 \chi_0}.
\end{equation}
The magnetic stored energy is given by
\begin{multline}
    \mathscr{W}^{\left(m\right)}_\text{stored} = \frac{\mu_0}{4} \int_V \left\| \mathbf{H}_h \right\|^2\dV + \frac{\mu_0}{4} \int_{\Vext} \left\| \mathbf{H}_h \right\|^2\dV \\=  \frac{1}{4\mu_0} \int_V \left\| \nabla \times \mathbf{A}_h \right\|^2\dV + \frac{1}{4\mu_0} \int_{\Vext} \left\| \nabla \times \mathbf{A}_h \right\|^2\dV
\end{multline}
where $\mathbf{A}_h= \mu_0 \frac{\ell_c^2}{\gammato{h}}\mathbf{j}_h^\perp$ is the vector potential associated with the magnetoquasistatic current mode. By using the identity
\begin{equation}
    \nabla \times \mathbf{A} \cdot \nabla \times \mathbf{B} = \mathbf{A} \cdot \nabla \times   \nabla \times \mathbf{B} + \nabla  \cdot \left[ \mathbf{A} \times  \nabla \times \mathbf{B} \right]
\end{equation}
and the property \cite{forestiere_magnetoquasistatic_2020} $\nabla \times   \nabla \times \mathbf{A}_h =\left({\chi^\perp_h}/{\ell_c^2}\right) \, \mathbf{A}_h ,\,\, \mathbf{r} \in V$,
% \begin{align}
%      \nabla \times   \nabla \times \mathbf{A}_h &= \left\{
%      \begin{array}{cc}
%         \left({\chi^\perp_h}/{\ell_c^2}\right) \, \mathbf{A}_h  & \mathbf{r} \in V \\
%          \mathbf{0} & \mathbf{r} \notin V
%      \end{array}\right.
% \end{align}
we obtain
\begin{equation}
    \mathscr{W}_\text{stored}^{\left(m\right)} = \frac{1}{4\mu_0} \int_{V } \mathbf{A}_h \cdot \nabla \times \nabla \times \mathbf{A}_h\dV =  \frac{\mu_0 \ell_c^2}{4 \chi^\perp_h} \left\| \mathbf{j}_h^\perp \right\|^2.
\end{equation}
Thus, the total stored energy $\mathscr{W}_\text{stored}$ is
\begin{equation}
      \mathscr{W}_\text{stored} = \mathscr{W}_\text{stored}^{\left(e\right)} +   \mathscr{W}_\text{stored}^{\left(m\right)} =  \frac{\mu_0}{2} \frac{\ell_c^2}{\chi^\perp_h} \left\| \mathbf{j}_h^\perp \right\|^2.
\end{equation}
By using the following identity:
\begin{multline}
 \frac{\mu_0}{2} \frac{\ell_c^2}{\gammato{h}} =  \frac{1}{\| \mathbf{j}_h^\perp \|^2}   \frac{\mu_0}{8 \pi} \int_{V} \Jto_h  \rp \cdot \int_{V}  \frac{\Jto_h \rpp}{\left| {\bf r} - {\bf r}' \right|} \dV' \dV
\end{multline}
the expression of total stored energy is rewritten as:
\begin{multline}
      \mathscr{W}_\text{stored} = \mathscr{W}_\text{stored}^{\left(e\right)} +   \mathscr{W}_\text{stored}^{\left(m\right)} =  \frac{\mu_0}{2} \frac{\ell_c^2}{\chi^\perp_h} \left\| \mathbf{j}_h^\perp \right\|^2 = \\ \frac{\mu_0}{8 \pi} \int_{V} \Jto_h  \rp \cdot \int_{V}  \frac{\Jto_h \rpp}{\left| {\bf r} - {\bf r}' \right|} \dV' \dV.
      \label{eq:WstoredM}
\end{multline}
%We note that stored energy of an electrically small high permittivity resonator is two times the expression of the stored magnetic energy obtained Vandenbosch \cite{vandenbosch_reactive_2010} for a non-dispersive PEC radiator of the magnetic-type.

{\bf Radiated Power -}
The power $\mathscr{P}_m$ radiated by the polarization current mode $\mathbf{j}_h^\perp$ with non-vanishing magnetic dipole moment is given by
\begin{multline}
   \mathscr{P}_m = \frac{\omega_h^4/c_0^3}{12 c_0 \pi} \left\| \mathbf{M}_h \right\|^2 =  \omega_h \frac{\mu_0}{12 \pi} \left( \frac{\omega_h}{c_0} \ell_c \right)^3 \frac{1}{\ell_c^3}\left\| \mathbf{M}_h \right\|^2 = \\ \omega_h \frac{\mu_0  }{48 \pi} \left( k_0 \ell_c \right)^3 \frac{1}{\ell_c^3} \int_{V} \Jto_h \rp \cdot \int_{V} \Jto_h \rpp \left| {\bf r} - {\bf r}' \right|^2 \dV \dV'.
   \label{eq:PowerM}
\end{multline}

{\bf Q factor -}
We now compute the Q factor by using the definition . We assume that the material losses are negligible; thus all the contribution to the power loss comes from the power radiated to infinity $\mathscr{P}_m$. By combining Eqs. \ref{eq:WstoredM} and  \ref{eq:PowerM} and the definition \ref{eq:QfacDef} we obtain Eq. \ref{eq:QfactorDiel}. It is worth noting that the expression \ref{eq:QfactorDiel} coincides with the one derived by Vandenbosch in Ref. \cite{vandenbosch_simple_2011} for an electrically small tuned PEC radiator of the magnetic-type.

\subsection{Resonant modes of surface scatterers}
\label{sec:SurfRes}

\subsubsection{Quasi-electrostatic resonances}
\label{sec:SurfResEQS}
Resonant  electromagnetic  scattering  from a small-size non-magnetic scatterer occupying the surface $S$ may occur when  the  imaginary  part  of its surface conductivity $\surfcond$ is negative \cite{forestiere_electromagnetic_2019}. The corresponding quasi-electrostatic surface current modes are solution of the eigenvalue problem
\begin{equation}
     {\bf j}_h^\parallel \rp  =  \gammalo{h} \ell_c \, \hat{\mathbf{n}} \times \hat{\mathbf{n}} \times \tnabla_S \oint_{S}     \frac{ \tnabla_{S'} \cdot  {\bf j}_h^\parallel \left( {\bf r}' \right)}{ 4 \pi \left| \rbt - \rbt' \right| }
       \dS' \qquad \forall \rt \in S,
       \label{eq:3DplasmonsSurf}
\end{equation}
where $\nabla_S$ is the surface gradient and $\nabla_S'\cdot$ is the surface divergence. All considerations made for the three dimensional scatterers can be transplanted in this case, considering the scalar product $\displaystyle \langle  \mathbf{f} , \mathbf{g}  \rangle_{S} = \int_{S}  \mathbf{f}^*   \cdot  \mathbf{g}\,   \dS$.
% \begin{equation}
%   \langle  \mathbf{f} , \mathbf{g}  \rangle_{S} = \int_{S}  \mathbf{f}^*   \cdot  \mathbf{g}\,   \dS.
%  \label{eq:ScalarProd2}
% \end{equation}
The surface current density field $\mathbf{J}^s\rp$ induced on the scatterer by an incident electric field ${\bf E}_{inc}$ is given by \cite{forestiere_electromagnetic_2019}
\begin{equation}
    \mathbf{J}^s \left( \mathbf{r} \right) \approx  \sum_{h} \frac{ \surfcond \left( \omega \right) \xi}{\xi +  \gammalo{h} \surfcond \left( \omega \right) } \langle   \mathbf{j}_h^\parallel, {\bf E}_{inc} \rangle_{S} \, \mathbf{j}_h^\parallel \rp \; 
\end{equation}
where $\surfcond$ is the surface conductivity of the scatterer.
The resonance frequency $\omega_h$ of the $h$-th quasi-electrostatic current mode is the frequency at which the real part of the denominator in the above equation vanishes \cite{forestiere_electromagnetic_2019}
\begin{equation}
    \text{Im} \left\{ \surfcond \left( \omega_h \right) \right\} = -\frac{1}{ \gammalo{h}} \left( \frac{\omega_h}{c_0} \ell_c \right).
\end{equation}
The electric dipole moment $\PE{h}$ of the surface current mode $ \Jlo_{h} $ is obtained by Eq. \ref{eq:Dipole}, where the integration is now performed on the surface, while the Q factor is still given by Eq. \ref{eq:QfactorEQS}.

\subsubsection{Quasi-magnetostatic resonances}
\label{sec:SurfResMQS}
Resonant  electromagnetic  scattering  from  a  small-size non-magnetic  scatterer occupying the surface $S$  may occur when  the  imaginary  part  of its the surface  conductivity $\surfcond$  is positive and sufficiently high \cite{forestiere_electromagnetic_2019}. The corresponding quasi-magnetostatic resonances are associated with the eigenvalues of the integral operator that relates the vector potential to the surface current density:
\begin{equation}
\Jto_h \rp =-\frac{\gammato{h}}{\ell_c } \, \hat{\mathbf{n}} \times  \hat{\mathbf{n}} \times \int_{S} \frac{ \Jto_h \rpp}{4\pi \left| \rbt - \rbt' \right|} \dS'.
\label{eq:QMSproblemSurf}
\end{equation}
All considerations made for three-dimensional scatterers can then be transplanted to this scenario. If a surface with surface conductivity $\surfcond$ is excited by an incident electric field $\mathbf{E}_{inc}$, the surface current density field $\mathbf{J}^s\rp$ induced on the surface is given by \cite{forestiere_electromagnetic_2019}
\begin{equation}
    \mathbf{J}^s \left( \mathbf{r} \right) \approx  \sum_{h} \left(\frac{ \Sigma \left( \omega \right)}{-1 +  \xi\gammato{h} \left( \xi \right) \Sigma \left( \omega \right) }\right) \langle   \mathbf{j}_h^\perp, {\bf E}_{inc} \rangle_{S} \, \mathbf{j}_h^\perp \rp.
\end{equation}
The quasi-magnetostatic resonance frequency $\omega_h$ of the $h$-th mode is the frequency at which \cite{forestiere_electromagnetic_2019}
\begin{equation}
    \text{Im} \left\{ \Sigma \left( \omega_h \right) \right\} = \frac{1}{ \gammato{h} } \frac{c_0}{\omega_h \ell_c}.
\end{equation}
The corresponding size parameter at the resonance is given by \ref{eq:SizeParameter}. The magnetic dipole moment $\PM{h}$ of the quasi-magnetostatic current mode $ \Jto_{h} $ is obtained by Eq. \ref{eq:DipoleM}, where the integration is now performed on the surface $\tSigma$, while the Q factor is given by Eq. \ref{eq:QfactorDiel}.

\section*{Acknowledgment}
This work was partially supported by the AFOSR MURI program and the Simons Foundation.

\end{document}